\documentclass[12pt]{article}
\usepackage[colorlinks,linkcolor=Blue,citecolor=Blue,bookmarks,bookmarksnumbered]{hyperref}
\usepackage[scaled=0.85]{helvet}
\usepackage{XoohmE,accents}
\usepackage{graphicx,color}
\usepackage{booktabs}
\usepackage{multirow}
\usepackage{placeins} 

\newcommand{\bm}[1]{\mbox{\boldmath$#1$}}

\def\be{\begin{equation}}
\def\ee{\end{equation}}

\def\fracm#1#2{\hbox{\large{${\frac{{#1}}{{#2}}}$}}}

\def\rI{{{}_{\rm I}}}
\def\rJ{{{}_{\rm J}}}

\def\hj{{\hat\jmath}}
\def\hk{{\hat k}}
\def\hi{{\hat\imath}}

\def\Tilde#1{\widetilde{#1}}                    
\def\Hat#1{\widehat{#1}}                        

\def\DDt#1{\accentset{\hbox{\LARGE.\kern-2pt.}}{#1}}	
\def\ddt#1{\accentset{\hbox{\large\kern.5pt.\kern-1pt.}}{#1}}	
\def\eX{\rlap{\raisebox{.35ex}{\kern.45ex\scriptsize\it=}}{\boldsymbol X}}
\def\eY{\rlap{\raisebox{.35ex}{\kern.12ex\scriptsize\it=}}{\boldsymbol Y}}
\def\EX{\rlap{\raisebox{.45ex}{\kern.425ex\scriptsize\it=}}{X}}
\def\EY{\rlap{\raisebox{.45ex}{\kern.1ex\scriptsize\it=}}{Y}}

\def\hj{{\hat\jmath}}

\def\bDb{\hbox{\kern2pt\vrule height10pt depth-9.2pt width6pt\kern-9pt{$\boldsymbol D$}}\mkern-2mu}

\def\bQb{\hbox{\kern2pt\vrule height10pt depth-9.2pt width6pt\kern-9pt{$\boldsymbol Q$}}}

\def\BSb{\hbox{\kern2.5pt\vrule height10pt depth-9.2pt width7pt\kern-10.25pt{$\boldsymbol{\mit\Sigma}$}}}

\def\rQb{\hbox{\kern1pt\vrule height10pt depth-9.2pt width6pt\kern-8pt{\bf Q}}}

\def\rBx#1#2{\hbox to#1{#2\hss}}

\definecolor{Hey}{rgb}{.9,.05,.4}
\definecolor{plum}{rgb}{.4,0,.6}

\definecolor{Green}  {rgb}{0.10,0.70,0.10} 
\definecolor{Orange} {rgb}{1.00,0.50,0.15} 
\definecolor{Red}    {rgb}{0.90,0.00,0.12} 
\definecolor{Purple} {rgb}{0.50,0.25,0.55} 
\definecolor{Turque} {rgb}{0.00,0.65,0.85} 
\definecolor{Blue}   {rgb}{0.00,0.00,1.00} 
\definecolor{Magenta}{rgb}{1.00,0.00,1.00} 
\definecolor{Gold}   {rgb}{1.00,0.75,0.25} 
\definecolor{Seaweed}{rgb}{0.01,0.24,0.09} 
\definecolor{Brown}  {rgb}{0.43,0.26,0.32} 
\definecolor{grey1}  {rgb}{0.20,0.20,0.20} 
\definecolor{grey2}  {rgb}{0.40,0.40,0.40} 
\definecolor{grey3}  {rgb}{0.60,0.60,0.60} 
\definecolor{grey4}  {rgb}{0.80,0.80,0.80} 
\definecolor{grey5}  {rgb}{0.90,0.90,0.90} 
\def\C#1#2{{\ifcase#1\or
             \color{Green}\or \color{Orange}\or \color{Red}\or
              \color{Purple}\or \color{Turque}\or \color{Blue}\or
               \color{Magenta}\or \color{Gold}\or \color{Seaweed}\or
                \color{Brown}\or\color{grey1}\or\color{grey2}\or
                 \color{grey3}\else\color{grey4}\fi#2}}

\definecolor{Slate} {rgb}{0.00,0.45,0.55}

\def\Ft#1{\,\footnote{#1}}
\newdimen\parshift\parshift=\parindent
\catcode`@=11
 \long\def\@footnotetext#1{\insert\footins{\reset@font\footnotesize
           \interlinepenalty\interfootnotelinepenalty\splittopskip%
            \footnotesep\splitmaxdepth\dp\strutbox\floatingpenalty\@MM%
             \hsize\columnwidth\addtolength{\hsize}{-2\parindent}
              \@parboxrestore\protected@edef\@currentlabel%
              {\csname p@footnote\endcsname\@thefnmark}%
                \color@begingroup%
                 \@makefntext{\rule\z@\footnotesep\ignorespaces#1%
                  \@finalstrut\strutbox}%
                \color@endgroup}}
 \long\def\@makefntext#1{\hglue\parshift%
           \vbox{\noindent\baselineskip=11pt plus.5pt minus.5pt\hb@xt@0em{\hss\@makefnmark\kern1pt}#1}}
\catcode`@=12
 \font\rOpe=cmsy10                        
 \def\ktl{{\hbox{\rOpe\char'170}}}        
 \def\kbl{{\hbox{\rOpe\char'170}}}        
 \def\kcr{{\reflectbox{\rOpe\char'170}}}        
 \def\ktr{{\reflectbox{\rOpe\char'170}}}        
 \def\kbr{{\reflectbox{\rOpe\char'170}}}        
 \def\Border{\vbox{\hsize0pt
        \setlength{\unitlength}{1mm}
        \newcount\xco
        \newcount\yco
        \xco=-21
        \yco=12
        \begin{picture}(0,0)(-7.5,0)
        \put(\xco,\yco){$\ktl$}
        \advance\yco by-1
        {\loop
        \put(\xco,\yco){$\kcr$}
        \advance\yco by-2
        \ifnum\yco>-240
        \repeat
        \put(\xco,\yco){$\kbl$}}
        \xco=170
        \yco=12
        \put(\xco,\yco){$\ktr$}
        \advance\yco by-1
        {\loop
        \put(\xco,\yco){$\kcr$}
        \advance\yco by-2
        \ifnum\yco>-240
        \repeat
        \put(\xco,\yco){$\kbr$}}
        \put(-19.5,13){\scalebox{.6065}{%
         University of Maryland Center for String and Particle  Theory \&\ Physics Department%
        |University of Maryland Center for String and Particle  Theory \&\ Physics Department}}
        \put(-19.5,-241.5){\scalebox{.5835}{%
         Howard University Department of Physics and Astronomy%
        |Howard University Department of Physics and Astronomy%
        |Howard University Department of Physics and Astronomy}}
        \end{picture}
        \par\vskip-8mm}}
\definecolor{UMred}{rgb}{.9,.05,.2}
\definecolor{HUblue}{rgb}{.0,.3,.7}
 \def\UMbanner{\vbox{\hsize0pt
        \setlength{\unitlength}{.4mm}
        \thicklines\color{UMred}
        \begin{picture}(0,0)(-30,-10)
        \put(165,16){\line(1,0){4}}
        \put(170,16){\line(1,0){4}}
        \put(180,16){\line(1,0){4}}
        \put(175,0){\line(1,0){4}}
        \put(180,0){\line(1,0){4}}
        \put(185,0){\line(1,0){4}}
        \put(169,0){\line(0,1){16}}
        \put(170,0){\line(0,1){16}}
        \put(179,0){\line(0,1){16}}
        \put(180,0){\line(0,1){16}}
        \put(184,0){\line(0,1){16}}
        \put(185,0){\line(0,1){16}}
        \put(169,16){\oval(8,32)[bl]}
        \put(170,16){\oval(8,32)[br]}
        \put(179,0){\oval(8,32)[tl]}
        \put(185,0){\oval(8,32)[tr]}
        \color{HUblue}
        \put(167.75,-2){\line(1,0){4}}
        \put(172.75,-2){\line(1,0){4}}
        \put(177.75,-2){\line(1,0){4}}
        \put(182.75,-2){\line(1,0){4}}
        \put(167.75,-2){\line(0,-1){16}}
        \put(171.75,-2){\line(0,-1){16}}
        \put(172.75,-2){\line(0,-1){16}}
        \put(176.75,-2){\line(0,-1){16}}
        \put(181.75,-2){\line(0,-1){16}}
        \put(182.75,-2){\line(0,-1){16}}
        \put(181.75,-2){\oval(8,32)[bl]}
        \put(182.75,-2){\oval(8,32)[br]}
        \put(167.75,-18){\line(1,0){4}}
        \put(172.75,-18){\line(1,0){4}}
        \end{picture}
        \par\vskip-6.5mm
        \thicklines}}

 \SfTitles 
 \seceq

\begin{document}

\thispagestyle{empty}
\vbox{\Border\UMbanner}
 \noindent{\small
 \today\hfill{PP 012-019 
 }}
  \vspace*{5mm}
 \begin{center}
{\LARGE\sf\bfseries 
 Adinkras and SUSY Holography: \\ 
\vskip.1in  Some Explicit Examples
}\\[12mm]
{\large\sf\bfseries S.\ James Gates, Jr.$^*$\footnote{gatess@wam.umd.edu},~
                    T.\, H\"{u}bsch$^{\dag\ddag}$\footnote{thubsch@howard.edu},~ and\, 
                    Kory Stiffler$^*$\footnote{kstiffle@umd.edu}
                    }\\*[4mm]
\emph{
      \centering
      $^*$Center for String and Particle Theory, Dept.\ of Physics, \\ University of Maryland, College Park, MD 20472,  
   \\\vspace{1.0mm}
     $^\ddag$Dept.\ of Physics, University of Central Florida, Orlando, FL 
     \\\vspace{1.0mm}
      $^\dag$Dept.\ of Physics \&\ Astronomy, Howard University, Washington, DC 20059
}
 \\\vspace*{3.6in}
{\sf\bfseries ABSTRACT}\\[4mm]
\parbox{150mm}{\parindent=2pc\indent\baselineskip=13pt plus1pt
We discuss the mechanism by which adinkras holographically store the
required information for the $\textit{Spin}(1,3)$ Clifford Algebra fiber bundle in the
cases of three 4D, ${\cal N}=1$ representations: the chiral, vector and tensor
supermultiplets.
 }
 
\end{center}

%

\newpage

\section{Introduction}  

$~~~~$ Some time ago, {\cite {G1}} an effort was made to extract the relationships between some 4D,
${\cal N}=1$ ($N=4$) supermultiplets and one of the ${\cal GR}(\rd, N)$ Algebras\Ft{The label ``${\cal GR}(\rd,N)$'' specifies the algebra generated by $N$ L- and $N$ R-matrices of size $\rd\,{\times}\,\rd$, akin to van der Waerden's $\s$- and $\bar\s$-matrices; see equations~(\ref{GarDNAlg1})--(\ref{GarDNAlg2}), below. ${\cal N}$ counts supersymmetry in terms of the smallest spinors in the given (space)time, while $N$ counts the individual, real supercharge components. Finally, (space)time dimension is denoted by a capital D, such as in ``1D'' {\em\/vs.\/}\ ``4D.''}
 (or `Garden Algebras') 
that arise in an approach to understanding the representations of one dimensional supersymmetrical quantum 
mechanical systems.  These algebras \cite{GRana}, generated by L- and R-matrices, are extended real versions of the algebras that arise from the van der Waerden formalism.
It has been proposed that these can be taken as the basic 
building blocks of a rigorous theory of {\em {off}}-{\em {shell}} representations for space-time SUSY
in much the same way as quark triplets and anti-triplets are the basic building blocks for hadrons.  For  4D, ${\cal N}=1$ supermultiplets, the appropriate Garden Algebra has $\rd=4$ and $N=4$.  The proposal 
that the same sets of mathematical objects provide the foundation for {\em {both}} supersymmetrical 
quantum mechanical systems {\em {and}} higher dimensional supersymmetrical quantum field theories 
implies the possible existence of a type of holography, given the name `SUSY Holography' in the 
work of {\cite {ENUF}}.

An unexpected and empowering development in this approach was the realization that  Garden 
Algebras can be obtained from graph theory {\cite {adinkra1}}.  The required graphs have been 
given the names of `adinkras' and these are adept at capturing an attribute (though present in
all such descriptions) of these systems that is less conveniently described in other approaches.  
This attribute has been given the name `height' and corresponds to the engineering dimension of 
the fields that occur in a supersymmetrical representation.  Other contributions
to the development of this approach that have been facilitated by adinkras are:
\par\smallskip
\begin{minipage}{160mm}
\begin{itemize}\vspace{-3mm}\itemsep=-3pt\raggedright
  \item[(a.)] the discovery that adinkras (and therefore off-shell supersymmetry representation 
theory) describe spaces of `marked cubical topology' \cite{Adnktop}, 
  \item[(b.)] the irreducible representations of adinkras are determined by \\ self-dual block
linear error-correcting codes \cite{Codes}, and
 \item[(c.)] that there exist a relation to mathematical structures called
 posets \cite{Posets}.
\end{itemize}
\end{minipage}

The L-matrices have a well defined role for one dimensional quantum mechanical systems.  But 
how can these structures be used to re-construct a Dirac-operator for a higher 
dimensional field theories with supersymmetry?  If the `SUSY Holography' 
conjecture in the work of {\cite {ENUF}} is correct, there must be a way to 
systematically achieve this goal.  In fact, Ref.~\cite{dimEnhanc} contains explicit demonstrations 
along these lines for some specific examples and gives a general requirement of necessary conditions to achieve SUSY 
holography between 1D, ${\cal N}=4$ models and 4D, ${\cal N}=1$ systems.
At least in the methods of implementation shown in  Ref.~\cite{dimEnhanc} the
heights of nodes plays a role.  A rather different approach \cite{Bowtie}
has been achieved for relating 1D $N$-extend SUSY models and 2D, $(p,q)$-supersymmetric
systems.  In this approach it has been shown that there exist obstructions, in the form of
minimal length closed co-cycles, whose absence permit the `liftability' of the 
data described by the L-matrices to successfully describe 2D models.

In the case of 4D, ${\cal N}=1$ SUSY it turns out that there is sufficient information in the 
work of {\cite {G1}} to see how, by beginning from a four dimensional field theory with 
simple supersymmetry, one can uncover a  relation to the L-matrices of a 1D, $\cal N$ 
$=$ 4 formalism.  It is a more intricate problem to see how one can begin {\em {solely}} 
within a one dimensional `Plato's Cave' \cite{PCave} and then in a systematic manner `discover' 
that all the required data is already present within those confines.  This will be treated
in a separate publication.  So the primary purpose of this work is to give an
explicit discussion based on the results in \cite{G1} showing how the data 
associated with adinkras is embedded with the structure of some dimensional
reduced d $=$  4, $N=4$ Garden Algebras, ${\cal GR}(4,4)$.

\section{Spinors Outside of Plato's Cave}  

$~~~~$ For the four dimensional physicist, the beginning of describing spinors can
start with the introduction of a set of Dirac gamma matrices.  Let us be even more
restrictive in our starting point by requiring that all four of the Dirac gamma matrices are
real.  It is simple to see that such a set is provided by ${\bm {\g}}^{\mu}$ $=$
$(\, {\bm {\g}}^{0} , \, {\bm {\g}}^{1} , \, {\bm {\g}}^{2} , \, {\bm {\g}}^{3}  \,)$ where
\be
 \eqalign{
&~~~~~~~~~ {(\gamma^0)}{}_a{}^b  = i ( \sigma^3
 \otimes \sigma^2  ){}_a{}^b 
~~~~,~~~~~~ {(\gamma^1)}{}_a{}^b  = ({\rm I}_2 
\otimes \sigma^1 ){}_a{}^b ~~~~~, \cr
&~~~~~~~~~ {(\gamma^2)}{}_a{}^b  = (\sigma^2 
\otimes \sigma^2 ){}_a{}^b ~~~~~,~~~~~~ 
  {(\gamma^3)}{}_a{}^b  = ({\rm I}_2 
\otimes \sigma^3  ){}_a{}^b  ~~~~~,
} 
\label{GammaM1}
\ee
written in terms of the outer product of the usual 2 $\times$ 2 Pauli matrices ($\bm {{\s}^1}$,  $\bm {{\s}^2}$,
$\bm {{\s}^3}$) and the 2 $\times$ 2 identity matrix $ {\bm {\rm I}}_2$.  These clearly satisfy
the usual Dirac condition
\be
{\bm {\g}}^{\mu} \, {\bm {\g}}^{\nu} ~+~ {\bm {\g}}^{\nu} \, {\bm {\g}}^{\mu}  ~=~ 2\, \eta^{\m \, \n}
\, {\bf I}_4 
\label{GammaM2}
\ee
where the 4 $\times$ 4 identity matrix ${\bf I}_4$ can also be written in terms of the outer
product,
\be
{\bf I}_4 ~=~   {\bm {\rm I}}_2 \otimes {\bm {\rm I}}_2   ~~~.  
\label{GammaM3}
\ee
Given a set of Dirac gamma matrices ${\bm {\g}}^{\mu}$, we can multiple higher powers of 
them and in particular define the usual product of four distinct ones to define ${\bm {\g}}^{5}$
via the usual definition
\be
 {\bm {\g}}^{\mu}  ~=~ i \, {\bm {\g}}^{0}  {\bm {\g}}^{1}  {\bm {\g}}^{2}  {\bm {\g}}^{3}
 \label{GammaM4}
\ee
which has the outer product representation
\be
 {(\gamma^5)}{}_a{}^b  = -(\sigma^1 \otimes \sigma^2 ){}_a{}^b   ~~~.  
 \label{GammaM5}
\ee
The three generators of spatial rotation acting on the spinor are provided by
${\bm \S}^{1 \, 2}$, ${\bm \S}^{2 \, 3}$, ${\bm \S}^{3 \, 1}$ where
where
\be
{\bm \S}^{1 \, 2} ~=~-\, i \, \fracm 12 \,  {\bm {\g}}^{1}  {\bm {\g}}^{2}  ~~,~~ 
{\bm \S}^{2 \, 3} ~=~- \, i \, \fracm 12 \,  {\bm {\g}}^{2}  {\bm {\g}}^{3}  ~~,~~ 
{\bm \S}^{3 \, 1} ~=~ - \, i \, \fracm 12 \,  {\bm {\g}}^{3}  {\bm {\g}}^{1}  ~~,~~ 
 \label{GammaM6}
\ee

At this point, a different set of 4 $\times$ 4 matrices can be introduced via 
the definitions
\be
\eqalign{
{~~} {\bm {\a}}^1  \,&=\, {\bm \s}^2 \otimes {\bm \s}^1   ~\,~~,~~
{\bm {\b}}^1   \,
=\, {\bm \s}^1 \otimes {\bm \s}^2 ~ \,\,, \cr
{\bm {\a}}^2  \,&=\, {\bf I}  \otimes {\bm \s}^2   ~~~~~~,~~
{\bm {\b}}^2  \,=\, {\bm \s}^2 \otimes {\bf I} ~~\, \,\,,   \cr  
{\bm {\a}}^3  \,&=\, {\bm \s}^2 \otimes {\bm \s}^3 ~\,~~,~~
{\bm {\b}}^3 \,=\,  {\bm \s}^3 \otimes {\bm \s}^2 \,\, \,\,,
}
\label{alpbet}
\ee
and it is a simple matter to show that the algebra for multiplying each of these 
is isomorphic to that of the usual 2 $\times$ 2 Pauli matrices.  Moreover, any 
element from the `$\bm \a$-set' commutes with any element from the `$\bm
\b$-set.'  This means that each of the two distinct sets can act as generators 
of an $SU(2)$ algebra.  The Dirac gamma matrices can be expressed using 
ordinary matrix multiplication of the `$\bm \a$-set' and `$\bm \b$-set' as
 \be
\eqalign{
{~~} {\bm {\g}}^0  \,&=\, i\, {\bm \b}^3   ~~\,~,~~~~
{\bm {\g}}^1   \,
=\, {\bm \a}^1  {\bm \b}^2    ~~,~~~~
{\bm {\g}}^2  \,=\, {\bm \a}^2  {\bm \b}^2  ~~\,\,~~,~~~~
{\bm {\g}}^3 \,=\,  {\bm \a}^3  {\bm \b}^2 
 \,\, \,\,.
}
\label{alpbet2}
\ee
which imply
\be
{\bm \S}^{1 \, 2} ~=~  \fracm 12 \,  {\bm {\a}}^{3}  ~~,~~ 
{\bm \S}^{2 \, 3} ~=~  \fracm 12 \,  {\bm {\a}}^{1}   ~~,~~ 
{\bm \S}^{3 \, 1} ~=~  \fracm 12 \,  {\bm {\a}}^{2}  ~~.
 \label{GammaM7}
\ee
and these equations make manifest that the $\bm \S$-matrices are the
generators of $SU(2)$.  But what has happened to the other $SU(2)$?

The definition of the ${\bm {\g}}^5 $ matrix implies
 \be
 {\bm {\g}}^5 ~=~ - \, {\bm {\b}}^1.
 \label{GammaM8}
\ee
and this together with the definition of  ${\bm {\g}}^0$ implies
 \be
\eqalign{
{~~} -i\,  \fracm 12 \, {\bm {\g}}^0  \,&=\, \fracm 12 \, {\bm \b}^3   ~~\,~,~~~~
-\, \fracm 12 \,{\bm {\g}}^5  \,=\,  \fracm 12 \, {\bm \b}^1    ~~,~~~~ \fracm 12 \,
{\bm {\g}}^0  {\bm {\g}}^5 \,=\, \fracm 12 \,{\bm \b}^2  ~~\,\,~~.
}
\label{GammaM9}
\ee
It is not commonly noted that given a set of four dimensional gamma matrices, it is
possible to use them to define {\em {two}} {\em {commuting}} $SU(2)$'s and this 
is independent of the representation chosen for the gamma matrices.  The name 
``extended R-symmetry" seems a reasonable moniker for the $SU(2)_\beta$, 
given that it is a group-theoretic ``extension" (in its precise technical sense) of 
the $U(1)$ symmetry  generated by $\gamma^5$, which is identifiable with the
``R-symmetry" or the ``axial symmetry." 

   In the work of \cite{G1}, this second less well-recognized $SU(2)$ symmetry plays an interesting role in the analysis of four dimensional supersymmetrical field theories when reduced to 1D.  When the theory is off-shell, both of these $SU(2)$ symmetries are realized on the anti-commutator algebra of the the supercharges.  When the theory is on-shell, only the $SU(2)$ symmetry associated with angular-momentum (rotations) is realized on the anti-commutator algebra of the supercharges.

\section{Some 4D, ~${\cal N}=1$ Supermultiplets Outside of Plato's Cave}

$~~~~$ For the four dimensional physicist knowledgeable about the simplest
4D, ${\cal N}=1$ representations, the chiral, vector, and tensor supermultiplets
can be respectively described by the sets of equations in (\ref{chi1}), (\ref{V1}) and
(\ref{ten1}).  For the superspace derivative D${}_a$ (equivalent to the supercharge) 
acting on the fields of the chiral supermultiplet $(A, \, B, \, \psi_a, \, 
F, \, G)$ we have,
\be \eqalign{
{\rm D}_a A ~&=~ \psi_a  ~~~, \cr
{\rm D}_a B ~&=~ i \, (\gamma^5){}_a{}^b \, \psi_b  ~~~, \cr
{\rm D}_a \psi_b ~&=~ i\, (\gamma^\mu){}_{a \,b}\,  \partial_\mu A 
~-~  (\gamma^5\gamma^\mu){}_{a \,b} \, \partial_\mu B ~-~ i \, C_{a\, b} 
\,F  ~+~  (\gamma^5){}_{ a \, b} G  ~~, \cr
{\rm D}_a F ~&=~  (\gamma^\mu){}_a{}^b \, \partial_\mu \, \psi_b   ~~~, \cr
{\rm D}_a G ~&=~ i \,(\gamma^5\gamma^\mu){}_a{}^b \, \partial_\mu \,  
\psi_b  ~~~.
} \label{chi1}
\ee
For the superspace derivative D${}_a$  acting on the fields of the vector
supermultiplet $(A_{\m}, \,  \l_a, \,  d)$ we have,
\be 
\eqalign{
{\rm D}_a \, A{}_{\mu} ~&=~  (\gamma_\mu){}_a {}^b \,  \l_b  ~~~, \cr
{\rm D}_a \l_b ~&=~   - \,i \, \fracm 14 ( [\, \gamma^{\mu}\, , \,  \gamma^{\nu} 
\,]){}_a{}_b \, (\,  \partial_\mu  \, A{}_{\nu}    ~-~  \partial_\nu \, A{}_{\mu}  \, )
~+~  (\gamma^5){}_{a \,b} \,    d ~~,  {~~~~~~} \cr
{\rm D}_a \, d ~&=~  i \, (\gamma^5\gamma^\mu){}_a {}^b \, 
\,  \partial_\mu \l_b  ~~~. \cr
} \label{V1}
\ee
Finally, for the superspace derivative D${}_a$  acting on the fields of the tensor
supermultiplet $(\varphi, \, B_{\m \, \n}, \,  \chi_a )$ we have,
\be
 \eqalign{
{\rm D}_a \varphi ~&=~ \chi_a  ~~~, \cr
{\rm D}_a B{}_{\mu \, \nu } ~&=~ -\, \fracm 14 ( [\, \gamma_{\mu}
\, , \,  \gamma_{\nu} \,]){}_a{}^b \, \chi_b  ~~~, \cr
{\rm D}_a \chi_b ~&=~ i\, (\gamma^\mu){}_{a \,b}\,  \partial_\mu \varphi 
~-~  (\gamma^5\gamma^\mu){}_{a \,b} \, \e{}_{\mu}{}^{\rho \, \s \, \t}
\partial_\rho B {}_{\s \, \t}~~. {~~~~~~~~~~~~~~~~~~~~~~}
} \label{ten1}
\ee

In the work of \cite{G1}, there was introduced the notion that a structure
similar to the Wigner `little group' exists for all higher dimensional SUSY
models.  The process begins by restricting coordinate dependence of
all field to {\em {only}} depend on the temporal direction and was given
the name of `reducing on a $0$-brane.'   For gauge fields, only the 
Coulomb gauge-fixed components are retained.  Finally, one performs the 
redefinitions
\be\label{e:Rsymm}
F ~\to ~ \partial_{\t} F ~~,~~  G ~\to ~ \partial_{\t} G ~~,~~  
d ~\to ~ \partial_{\t} d ~~.~~  
\ee
This step ensures that the set of all bosons remaining have the same
engineering dimensions.  All the fermions already possess the same
engineering dimensions though of course this is different from that
of the bosons.  By this means, one obtains a truncation of the original
four dimensional theories supermultiplets that have now been mapped
into a set of one dimensional supermultiplets.   The redefinitions above also have the property that it permits the
 re-defined $F$ and $G$, together with $B$, to form a triplet under
 the extended R-symmetry.  Thus there occurs an enhanced symmetry
 within the confines of the cave.

\section{Some 1D, ~${\cal N}=4$ Supermultiplets Inside of Plato's Cave}

$~~~~$ Using the allusion to Plato's Cave, here everything depends only on the
temporal coordinate.  All bosonic and fermionic functions only depend
on a single time-like coordinate.  This is the realm of supersymmetric 
quantum mechanics.

The `L-matrices' of the Garden Algebra approach are real and correspond to 
the linking numbers of a topological space and can be directly read off (via a 
set of `Feynman-like rules') from the corresponding adinkra.  The parameter $N
$ describes the number of equivalence classes of such linking numbers and 
the parameter d specifies the size of the d $\times$ d L-matrices.  By 
definition, the `L-matrices' of the Garden Algebra approach satisfy the conditions,
\be \eqalign{
 (\,{\rm L}_\rI\,)_i{}^\hj\>(\,{\rm R}_\rJ\,)_\hj{}^k + (\,{\rm L}_\rJ\,)_i{}^\hj\>(\,{\rm 
 R}_\rI\,)_\hj{}^k &= 2\,\d_{\rI\rJ}\,\d_i{}^k~~,\cr
 (\,{\rm R}_\rJ\,)_\hi{}^j\>(\, {\rm L}_\rI\,)_j{}^\hk + (\,{\rm R}_\rI\,)_\hi{}^j\>(\,{\rm 
 L}_\rJ\,)_j{}^\hk
  &= 2\,\d_{\rI\rJ}\,\d_\hi{}^\hk~~,
}  \label{GarDNAlg1}
 \ee
where
\be
~~~~(\,{\rm R}_\rI\,)_\hj{}^k\,\d_{ik} = (\,{\rm L}_\rI\,)_i{}^\hk\,\d_{\hj\hk}~~.
\label{GarDNAlg2}
\end{equation}
For a fixed size d, there are $2^{d -1} d!$ distinct matrices that can be used.
Given a set of such L-matrices, one can introduce d bosons $\Phi_i $ ($i = 1, \, 
\dots, \, d $) and d fermions $ \Psi_{\hat k}$ (${\hat k}= 1, \, 
\dots, \, d $)
along with $N$ superderivatives ${\rm D}{}_{{}_{\rm I}}$ (${\rm I} = 1, \, 
\dots, \, N $)
that satisfy the equations
\be
{\rm D}{}_{{}_{\rm I}} \Phi_i ~=~ i \, \left( {\rm L}{}_{{}_{\rm I}}\right) {}_{i \, {\hat k}}  \,  \Psi_{\hat k}
~~~,~~~
{\rm D}{}_{{}_{\rm I}} \Psi_{\hat k} ~=~  \left( {\rm R}{}_{{}_{\rm I}}\right) {}_{{\hat k} \, i}  \,
{{d ~} \over { d \t}} \, \Phi_{i}  ~~.
 \label{chiD0J}
\ee
The definitions in (\ref{GarDNAlg1}), (\ref{GarDNAlg2}), and (\ref{chiD0J}) will ensure
that these D = 1 bosons and fermions
form a representation of $N$-extended SUSY.  All the bosons in this supermultiplet share
have same engineering dimensions and all the fermions in this supermultiplet share
have same engineering dimensions, but the latter dimensions are distinct from the former.
We call such 1D representations `valises.'  Let us restrict ourselves to the case 
where $\rd=4$ and $N=4$.  For the theoretical physicist inside Plato's cave, these
three sets of equations may be taken as a starting point.  

\begin{figure}[!ht]
   \begin{center}
   \begin{picture}(120,50)
      \put(0,0){\includegraphics[width = 120 \unitlength]{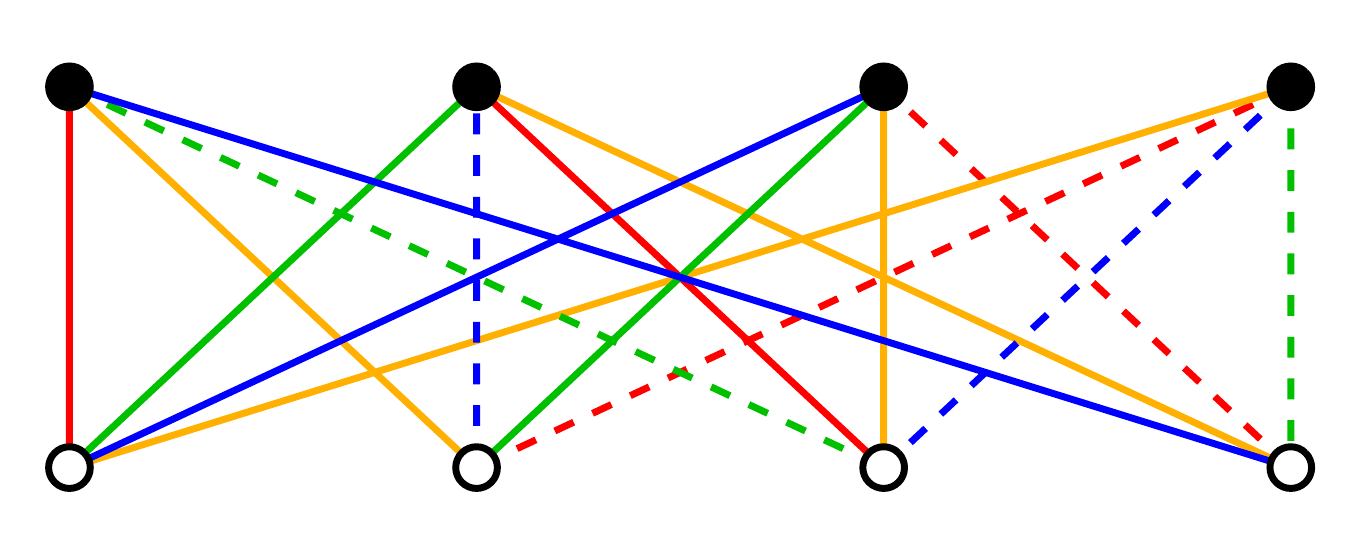}}
   \end{picture}
   \end{center}
   \caption{CM Adinkra.}
   \label{f:CM}
\end{figure}

\begin{figure}[!ht]
   \begin{center}
   \begin{picture}(120,50)
      \put(0,0){\includegraphics[width = 120 \unitlength]{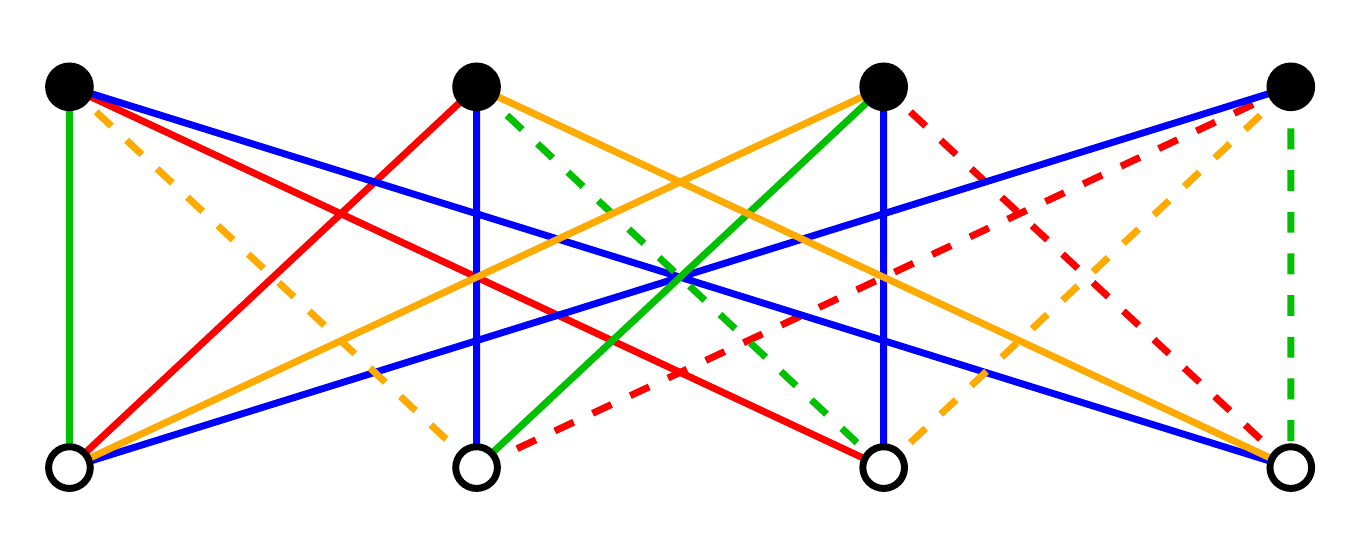}}
   \end{picture}
   \end{center}
   \caption{VM Adinkra.}
   \label{f:VM}
\end{figure}

\begin{figure}[!ht]
   \begin{center}
   \begin{picture}(120,50)
      \put(0,0){\includegraphics[width = 120 \unitlength]{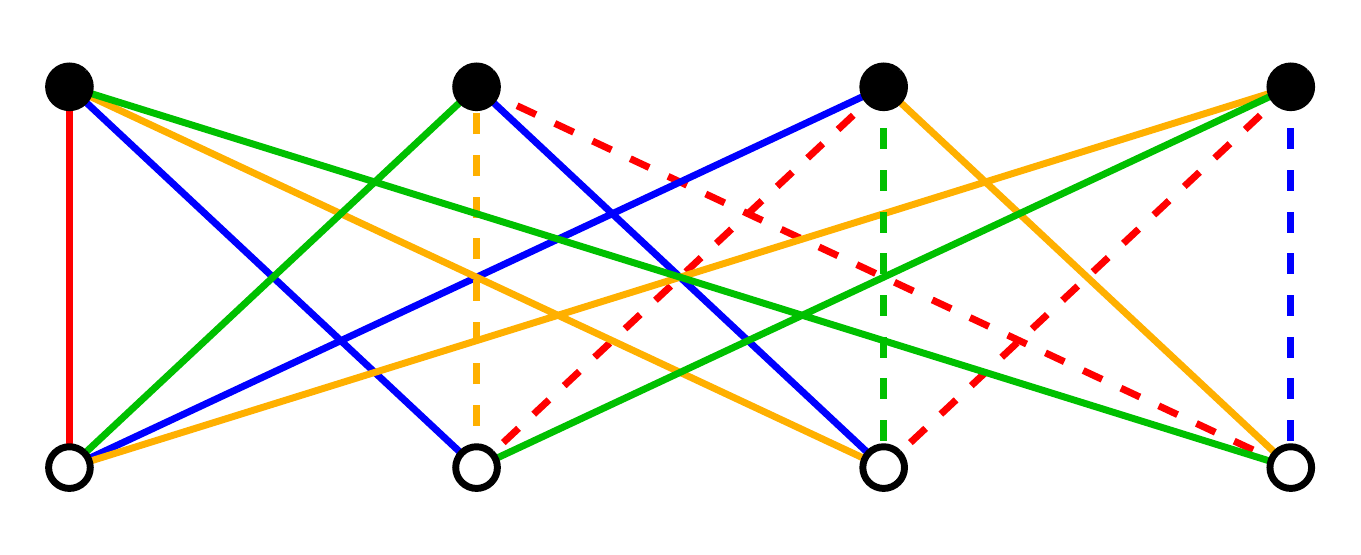}}
   \end{picture}
   \end{center}
   \caption{TM Adinkra.}
   \label{f:TM}
\end{figure}

Among the 1,536 quartets of L- and R-matrices that satisfy (\ref{GarDNAlg1}) and
(\ref{GarDNAlg2}) it would seem likely that some of 
these must coincide with the mathematical systems obtained in the previous chapter.  
A simple question to ask is, ``What set of L-matrices coincide so that the physicist
in Plato's cave is unknowingly describing the four dimensional multiplets when they
are reduced on a 0-brane?'' A priori, it is not at all obvious.  However, the work in
\cite{G1} has answered this question.  We know the reduced sets correspond to the
enunciations given in (\ref{chiD0F}), (\ref{V1D0E}), and (\ref{tenD0F}). These can also be displayed as the Adinkras in Figs.~\ref{f:CM}, \ref{f:VM}, and \ref{f:TM}.

CM~\hrulefill
$$
\left( {\rm L}{}_{1}\right) {}_{i \, {\hat k}}   ~=~
\left[\begin{array}{cccc}
~1 & ~~0 &  ~~0  &  ~~0 \\
~0 & ~~0 &  ~~0  &  ~-\, 1 \\
~0 & ~~1 &  ~~0  &  ~~0 \\
~0 & ~~0 &  ~-\, 1  &  ~~0 \\
\end{array}\right] ~~~,~~~
\left( {\rm L}{}_{2}\right) {}_{i \, {\hat k}}   ~=~
\left[\begin{array}{cccc}
~0 & ~~1 &  ~~0  &  ~ \, \, 0 \\
~0 & ~~ 0 &  ~~1  &  ~~0 \\
-\, 1 & ~~ 0 &  ~~0  &  ~~0 \\
~ 0 & ~~~0 &  ~~0  &   -\, 1 \\
\end{array}\right]  ~~~,
$$
\be
\left( {\rm L}{}_{3}\right) {}_{i \, {\hat k}}   ~=~
\left[\begin{array}{cccc}
~0 & ~~0 &  ~~1  &  ~~0 \\
~0 & ~- \, 1 &  ~~0  &  ~~0 \\
~0 & ~~0 &  ~~0  &  -\, 1 \\
~1 & ~~0 &  ~~0  &  ~~0 \\
\end{array}\right] ~~~,~~~
\left( {\rm L}{}_{4}\right) {}_{i \, {\hat k}}   ~=~
\left[\begin{array}{cccc}
~0 & ~~0 &  ~~0  &  ~ \, \, 1 \\
~1 & ~~ 0 &  ~~0  &  ~~0 \\
~0 & ~~ 0 &  ~~1  &  ~~0 \\
~ 0 & ~~~1 &  ~~0  &   ~~0  \\
\end{array}\right]  ~~.
 \label{chiD0F}
\ee

VM~\hrulefill
$$
\left( {\rm L}{}_{1}\right) {}_{i \, {\hat k}}   ~=~
\left[\begin{array}{cccc}
~0 & ~1 &  ~ 0  &  ~ 0 \\
~0 & ~0 &  ~0  &  -\,1 \\
~1 & ~0 &  ~ 0  &  ~0 \\
~0 & ~0 &  -\, 1  &  ~0 \\
\end{array}\right] ~~~,~~~
\left( {\rm L}{}_{2}\right) {}_{i \, {\hat k}}   ~=~
\left[\begin{array}{cccc}
~1 & ~ 0 &  ~0  &  ~ 0 \\
~0 & ~ 0 &  ~1  &  ~ 0 \\
 ~0 & - \, 1 &  ~0  &   ~ 0 \\
~0 & ~0 &  ~0  &  -\, 1 \\
\end{array}\right]  ~~~,
$$
\be {~~~~}
\left( {\rm L}{}_{3}\right) {}_{i \, {\hat k}}   ~=~
\left[\begin{array}{cccc}
~0 & ~0 &  ~ 0  &  ~ 1 \\
~0 & ~1 &  ~0  &   ~0 \\
~0 & ~0 &  ~ 1  &  ~0 \\
~1 & ~0 &  ~0  &  ~0 \\
\end{array}\right] ~~~~~~,~~~
\left( {\rm L}{}_{4}\right) {}_{i \, {\hat k}}   ~=~
\left[\begin{array}{cccc}
~0 & ~0 &  ~1  &  ~ 0 \\
-\,1 & ~ 0 &  ~0  &  ~ 0 \\
 ~0 & ~0 &  ~0  &   - \, 1 \\
~0 & ~1 &  ~0  &  ~  0 \\
\end{array}\right]  ~~~,
\label{V1D0E}
\ee

TM~\hrulefill
$$
\left( {\rm L}{}_{1}\right) {}_{i \, {\hat k}}   ~=~
\left[\begin{array}{cccc}
~1 & ~0 &  ~0  &  ~0 \\
~0 & ~0 &  -\, 1  &  ~ 0 \\
~0 & ~0 &  ~0  &  -\,1 \\
~0 & -\,1 &  ~ 0  &  ~0 \\
\end{array}\right] ~~~,~~~
\left( {\rm L}{}_{2}\right) {}_{i \, {\hat k}}   ~=~
\left[\begin{array}{cccc}
~0 & ~1 &  ~0  &  ~  0 \\
~0 & ~ 0 &  ~0  &  ~ 1 \\
~0 & ~ 0 &  -\,1  &  ~ 0 \\
 ~ 1 & ~0 &  ~0  &   ~ 0 \\
\end{array}\right]  ~~~,
$$
\be {~~~~}
\left( {\rm L}{}_{3}\right) {}_{i \, {\hat k}}   ~=~
\left[\begin{array}{cccc}
~0 & ~0 &  ~1  &  ~0 \\
~1 & ~0 &  ~ 0  &  ~ 0 \\
~0 & ~1 &  ~0  &   ~0 \\
~0 & ~0 &  ~ 0  &  -\, 1 \\
\end{array}\right] ~~~~~~,~~~
\left( {\rm L}{}_{4}\right) {}_{i \, {\hat k}}   ~=~
\left[\begin{array}{cccc}
~0 & ~0 &  ~0  &  ~  1 \\
~0 & -\, 1 &  ~0  &  ~ 0 \\
~1 & ~ 0 &  ~0  &  ~ 0 \\
 ~0 & ~0 &  ~1  &   ~ 0 \\
\end{array}\right]  ~~~,
\label{tenD0F}
\ee
$~~~~~~~~~$~\hrulefill

The next intriguing question to ask is, ``What remains of the $(1,3)$ Clifford Algebra
fiber bundle that allowed the spinors in four dimensions to be defined?''  Stated
in a different manner we can ask, ``Given that the physicist inside Plato's Cave
only has the data represented by the results in (\ref{chiD0F}), (\ref{V1D0E}), and
(\ref{tenD0F}), how is it possible to define the gamma matrices in (\ref{GammaM1}) knowing about
these three one dimensional $N$-extended supermultiplets?'' A prescriptive solution as a positive response to these questions is the essence of the ``SUSY
Holography Conjecture'' that was made in the work of \cite{ENUF}.

\section{How d $=4$, $N=4$ 
Adinkras Re-construct the Dirac Operator for 4D, ${\cal N}=1$ Supermultiplets Outside of Plato's Cave}

$~~~~$ The conditions in (\ref{GarDNAlg1}) and (\ref{GarDNAlg2}) define the Garden Algebra matrices.  The sets of
matrices for the (CM), (VM), and (TM) sets all can be shown to do satisfy these defining relations.  However, we can perform a
different calculation replacing the relative plus signs in the left-hand side of (\ref{GarDNAlg1}) by minus signs.  Here something interesting
occurs.  For each of the representations ($\cal R$) (given in (CM), (VM), and (TM)) one finds
there exists four sets of coefficients ${\k^{(\cal R)}}_{ \rI\rJ}{}^{\Hat a}$,
${{\Tilde \k}^{(\cal R)}}_{\rI\rJ}{}^{\Hat a}$, ${\ell^{(\cal R)}}_{ \rI\rJ}{}^{\Hat a}$,
${{\Tilde \ell}^{(\cal R)}}_{\rI\rJ}{}^{\Hat a}$ whose explicit values depend on the
representation used to perform the calculation. The coefficients appear in the 
equations,
\be \eqalign{
 (\,{\rm L}^{(\cal R)}_\rI\,)_i{}^\hj\>(\,{\rm R}^{(\cal R)}_\rJ\,)_\hj{}^k - (\,{\rm L}^{(\cal 
 R)}_\rJ\,)_i{}^\hj\>(\,{\rm R}^{(\cal R)}_\rI\,)_\hj{}^k
  &= i\, 2\,  \Big[ \, \k^{({\cal R})1}_{\rI\rJ}\, (\a_{1}){}_{i}{}^k
   +\k^{({\cal R})2}_{ \rI\rJ}\, (\a_{2}){}_{i}{}^k
   + \k^{({\cal R})3}_{\rI\rJ}\, (\a_{3}){}_{i}{}^k  \, \Big] \cr
 &~~ + i\, 2\, \Big[ \, {{\Tilde \k}^{(\cal R)}}_{\rI\rJ}{}^{1}\, (\b_{1}){}_{i}{}^k
  + {{\Tilde \k}^{(\cal R)}}_{\rI\rJ}{}^{2}\,(\b_{2}){}_{i}{}^k
  + {{\Tilde \k}^{(\cal R)}}_{\rI\rJ}{}^{3}\,(\b_{3}){}_{i}{}^k  \, \Big]  ~~,
}  \label{GarDNAlgK1}
\ee \vspace{1mm}
\be \eqalign{
 (\,{\rm R}^{(\cal R)}_\rI\,)_\hi{}^j\>(\, {\rm L}^{(\cal R)}_\rJ\,)_j{}^\hk - (\,{\rm R}^{(\cal 
 R)}_\rJ\,)_\hi{}^j\>(\,{\rm L}^{(\cal R)}_\rI\,)_j{}^\hk
  &= i\, 2\,  \Big[ \, \ell^{({\cal R})1}_{\rI\rJ}\, (\a_{1}){}_{\hat{i}}{}^{\hat{k}}
   +\ell^{({\cal R})2}_{ \rI\rJ}\, (\a_{2}){}_{\hat{i}}{}^{\hat{k}}
   + \ell^{({\cal R})3}_{\rI\rJ}\, (\a_{3}){}_{\hat{i}}{}^{\hat{k}}  \, \Big]
  \cr
 &~~+ i\, 2\,  \Big[ \, {{\Tilde \ell}^{(\cal R)}}_{\rI\rJ}{}^{1}\, (\b_{1}){}_{\hi}{}^{\hk} 
    +{{\Tilde \ell}^{(\cal R)}}_{\rI\rJ}{}^{2}\,(\b_{2}){}_{\hi}{}^{\hk}
    + {{\Tilde \ell}^{(\cal R)}}_{\rI\rJ}{}^{3}\,(\b_{3}){}_{\hi}{}^{\hk}  \, \Big]  ~~.
}  \label{GarDNAlgK2}
\ee
The six matrices that appear on the right hand of these equations are exactly the same
as the matrices denoted by the same symbols in equations in (\ref{alpbet}).  

Moreover, the matrices on the left hand side of (\ref{GarDNAlgK2}) from the view inside the cave, act
to map the space of the spinors back onto itself. We are thus motivated by (\ref{GammaM7}) and
(\ref{GammaM9}) to {\em {define}} relations of quantities inside the cave (the L-matrices) to quantities 
outside the cave (the $\gamma$-matrices) via the ``Adinkra/$\gamma$-matrix holography 
equation,''
\be
\eqalign{
  (\,{\rm R}^{(\cal R)}_\rI\,)_\hi{}^j\>(\, {\rm L}^{(\cal R)}_\rJ\,)_j{}^\hk - (\,{\rm R}^{(\cal 
 R)}_\rJ\,)_\hi{}^j\>(\,{\rm L}^{(\cal R)}_\rI\,)_j{}^\hk
  &= 2\Big[ \,\ell^{({\cal R})1}_{\rI\rJ}\, (\g^2 \g^3){}_{\hi}{}^\hk
   +  \ell^{({\cal R})2}_{\rI\rJ}\, (\g^3 \g^1){}_{\hi}{}^\hk
   + \ell^{({\cal R})3}_{\rI\rJ}\, (\g^1 \g^2){}_{\hi}{}^\hk   \, \Big] \cr
 &~- 2\Big[ \,  i\,  {{\Tilde \ell}^{(\cal R)}}_{
 \rI\rJ}{}^{1}\, (\g^5){}_{\hi}{}^\hk  \,-\,
  i\, {{\Tilde \ell}^{(\cal R)}}_{\rI\rJ}{}^{2}\, 
 (\g^0 \g^5){}_{\hi}{}^\hk  \,-\,  {{\Tilde \ell}^{(\cal R)}}_{\rI\rJ}{}^{3}\, 
 (\g^0){}_{\hi}{}^\hk 
  \, \Big].
}  \label{GarDNAdnk}
\ee
It should be noted that this is an over-constrained system of equations.  There are 
{\em a} {\em {prior}}i 96 independent conditions calculated on the LHS of this.  But 
the RHS asserts that these conditions are satisfied by only 36 parameters (i.\ e.\ the 
$\ell$ and $\Tilde \ell$ parameters).

Once this definition is accepted, it opens up a way for the physicist within the cave 
to construct a set of gamma matrices for the spinor bundle of the $(1,3)$ Clifford 
Algebra fiber bundle of a four dimensional spacetime.  Let us point out that one 
would {\em {not}} wish to use the equation in (\ref{GarDNAlgK1}) in this manner 
because there is no relevance to defining a set of $\g$-matrices that act on the 
space of the bosonic fields. Furthermore the second line on the RHS of \eq{GarDNAdnk} refers to 
the generators of the SU(2)-extended R-symmetry which is broken in the four 
dimensional theory (see the remarks near equation \eq{e:Rsymm}).

 By construction, the $\ell$, $\Tilde{\ell}$, $\k$, and $\Tilde{\k}$ coefficients satisfy $\ell_{{\rm I} {\rm J}} = - \ell_{{\rm J}{\rm I}}$, $\Tilde{\ell}_{{\rm I} {\rm J}} = - \Tilde{\ell}_{{\rm J}{\rm I}}$, $\k_{{\rm I} {\rm J}} = - \k_{{\rm J}{\rm I}}$, and $\Tilde{\k}_{{\rm I} {\rm J}} = - \Tilde{\k}_{{\rm J}{\rm I}}$. For the representations $CM$, $VM$, and $TM$, \emph{all} $\Tilde{\kappa}$ coefficients vanish and Eqs.~(\ref{e:ls}) and~\eq{e:ks} show all non-vanishing independent $\ell$, $\Tilde{\ell}$, and $\k$ coefficients.
\be
\begin{array}{cccccc}
 \ell _{12}^{(CM)2}=1 & \ell _{13}^{(CM)3}=1 & \ell _{14}^{(CM)1}=1 & \ell _{23}^{(CM)1}=1 & \ell _{24}^{(CM)3}=-1 & \ell
   _{34}^{(CM)2}=1 \\
 \Tilde{\ell }_{12}^{(VM)3}=-1 & \Tilde{\ell }_{13}^{(VM)2}=1 & \Tilde{\ell }_{14}^{(VM)1}=-1 & \Tilde{\ell }_{23}^{(VM)1}=1 &
   \Tilde{\ell }_{24}^{(VM)2}=1 & \Tilde{\ell }_{34}^{(VM)3}=1 \\
 \Tilde{\ell }_{12}^{(TM)3}=1 & \Tilde{\ell }_{13}^{(TM)2}=1 & \Tilde{\ell }_{14}^{(TM)1}=1 & \Tilde{\ell }_{23}^{(TM)1}=-1 &
   \Tilde{\ell }_{24}^{(TM)2}=1 & \Tilde{\ell }_{34}^{(TM)3}=-1
\end{array}
\label{e:ls}
\ee
\be
\begin{array}{cccccc}
 \kappa _{12}^{(CM)3}=-1 & \kappa _{13}^{(CM)1}=1 & \kappa _{14}^{(CM)2}=1 & \kappa _{23}^{(CM)2}=-1 & \kappa _{24}^{(CM)1}=1
   & \kappa _{34}^{(CM)3}=1 \\
    \kappa _{12}^{(VM)3}=-1 & \kappa _{13}^{(VM)2}=1 & \kappa _{14}^{(VM)1}=1 & \kappa _{23}^{(VM)1}=1 & \kappa _{24}^{(VM)2}=-1
   & \kappa _{34}^{(VM)3}=-1 \\
   \kappa _{12}^{(TM)1}=1 & \kappa _{13}^{(TM)2}=1 & \kappa _{14}^{(TM)3}=1 & \kappa _{23}^{(TM)3}=1 & \kappa _{24}^{(TM)2}=-1
   & \kappa _{34}^{(TM)1}=1
\end{array}
\label{e:ks}
\ee

\begin{figure}
\centering
    \begin{picture}(120,100)
    \put(0,0){\includegraphics[width = 120\unitlength]{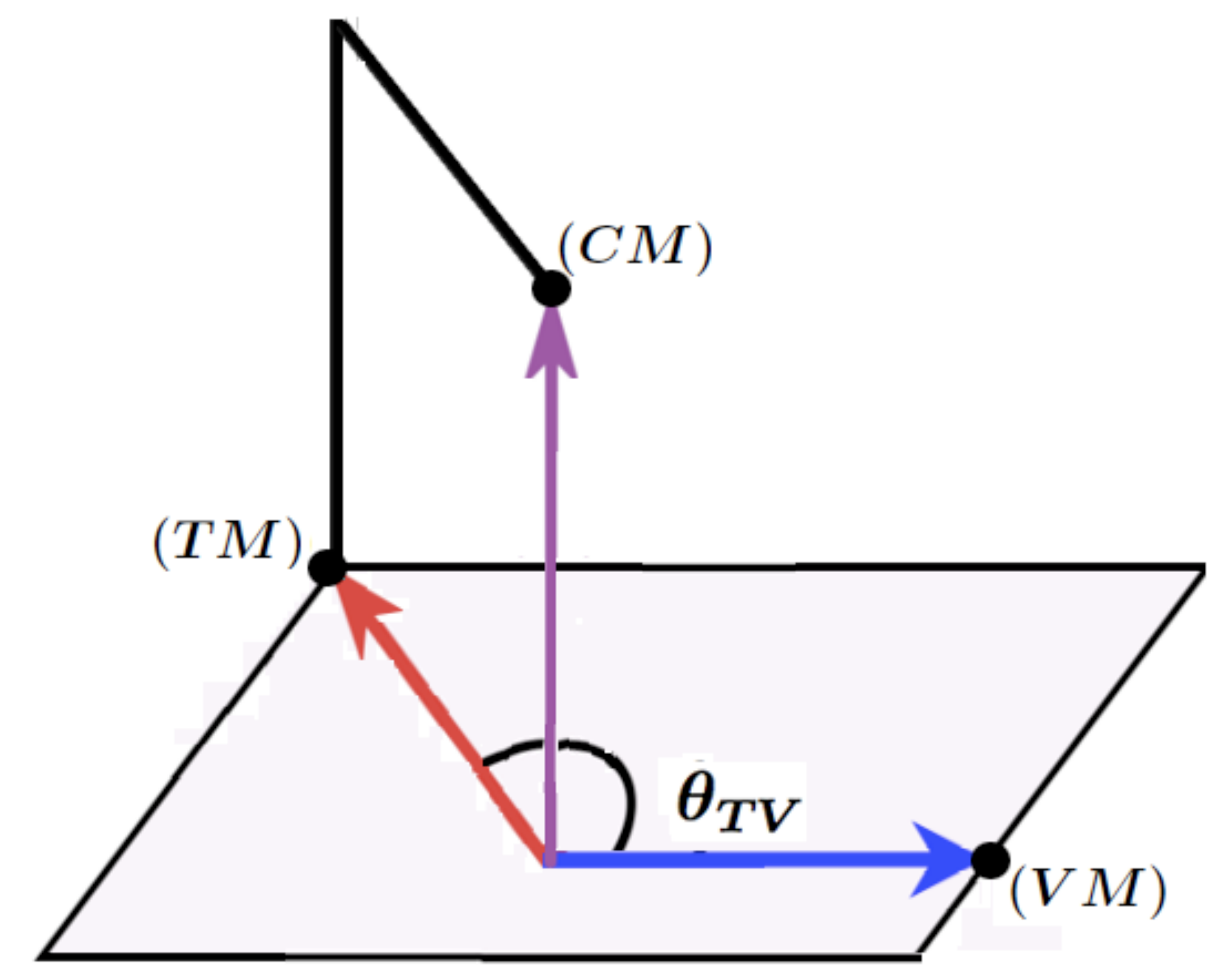}}
    \end{picture}
    \caption{A monoclinic lattice is built of two vectors (TM, VM) each perpendicular to a third (CM), but not perpendicular to each other.}
    \label{f:BoobTubeThetan}
\end{figure}

We see that these coefficients hold a rude surprise!
When the (CM) representation is used, {\em {all}} the $\Tilde{\ell}$-coefficients vanish.
When the (VM) and (TM) representations are used, {\em {all}} the $\ell$-coefficients 
vanish.  Thus, the cave dweller physicist is only able to reconstruct all of the gamma
matrices of the $(1,3)$ Clifford Algebra fiber bundle of a four dimensional spacetime
if and only if {\em {all}} the data from the (CM) representation and the data from at
least one of the (VM) or (TM) representations are used! Furthermore, independent of the representation, the $\ell$ coefficients satisfy 
\be\eqalign{
\ell_{{\rm I}{\rm J}}^{({\cal R}) \hat{a}} =&
\fracm 12 \, \e_{{\rm I}{\rm J}{\rm K}{\rm L}} \ell_{{\rm K}{\rm L
}}^{({\cal R})\hat{a}}   ~~,~~
 \Tilde{\ell}_{{\rm I}{\rm J}}^{({\cal R}) \hat{a}} =
- \, \fracm 12 \, \e_{{\rm I}{\rm J}{\rm K}{\rm L}}
\Tilde{\ell}_{{\rm K}{\rm L}}^{({\cal R})\hat{a}}
}
\ee
where as the $\k$ coefficients satisfy the basis dependent relations
\be\eqalign{
\k_{{\rm I}{\rm J}}^{(CM) \hat{a}} =&
-\fracm 12 \, \e_{{\rm I}{\rm J}{\rm K}{\rm L}} \k_{{\rm K}{\rm L}}^{(CM)\hat{a}} ~~,~~
\k_{{\rm I}{\rm J}}^{(VM) \hat{a}} = \fracm 12 \, \e_{{\rm I}{\rm J}{\rm K}{\rm L}} \k_{{\rm K}{\rm L}}^{(VM)\hat{a}} ~~,~~
\k_{{\rm I}{\rm J}}^{(TM) \hat{a}} = \fracm 12 \, \e_{{\rm I}{\rm J}{\rm K}{\rm L}} \k_{{\rm K}{\rm L}}^{(TM)\hat{a}}~.
}\ee
All of the $\k$, $\Tilde \k$, $\ell$, and $\Tilde \ell$ coefficients
 may be considered as the components of vectors in a
 seventy-two dimensional vector space.  Knowing this allows
 the physicist within the cave to recognize that the three
 supermultiplets defined by Eqs.~\eq{chiD0J}, \eq{chiD0F}, \eq{V1D0E}, and~\eq{tenD0F} are distinct
 representations. In the space of the $\k$, $\Tilde \k$, $\ell$, $\Tilde \ell$ parameters
 it is useful to introduce the notion of  inner product between the parameters
 of  representations.  If $\k^{({\cal R})}$, ${\Tilde \k}{}^{({\cal R})}$, $\ell{}^{({\cal R})}$,
 ${\Tilde \ell}{}^{({\cal R})}$ and $\k^{({\cal R}^{\prime})}$, ${\Tilde \k}{}^{({\cal R}^{
 \prime})}$, $\ell{}^{({\cal R}^{\prime})}$,
 ${\Tilde \ell}{}^{({\cal R}^{\prime})}$ are the parameters associated
 with the ${\cal R}$ and ${\cal R}^{\prime}$ representations, we can
 define
 \be \eqalign{
 \left[  ({\cal R}) \, {\bm {\cdot}} \,  ({\cal R}^{\prime}) \right]{}_{\k} &\equiv ~ \frc{1}{2}
\sum_{{\rm I}, {\rm J},  {\hat a} } \, \left[ ~ {\k}_{{\rm I}{\rm J}}^{\,
({\cal R}) \hat{a}} \,  {\k}_{{\rm I}{\rm J}}^{\, ({\cal R}^{\prime})
\hat{a}}  ~+~   \Tilde{\k}_{{\rm I}{\rm J}}^{\, ({\cal R}) \hat{a}}
\,  \Tilde{\k}_{{\rm I}{\rm J}}^{\, ({\cal R}^{\prime}) \hat{a}}  ~ \right]
~~~~, \cr
\left[  ({\cal R}) \, {\bm {\cdot}} \,  ({\cal R}^{\prime}) \right]{}_{\ell} ~&\equiv ~\frc{1}{2}
\sum_{{\rm I}, {\rm J},  {\hat a} } \, \left[ ~ {\ell}_{{\rm I}{\rm J}}^{\,
({\cal R}) \hat{a}} \,  {\ell}_{{\rm I}{\rm J}}^{\, ({\cal R}^{\prime})
\hat{a}}  ~+~   \Tilde{\ell}_{{\rm I}{\rm J}}^{\, ({\cal R}) \hat{a}}
\,  \Tilde{\ell}_{{\rm I}{\rm J}}^{\, ({\cal R}^{\prime}) \hat{a}}  ~ \right]
~~~.
}  \ee  
 The `angles' between the corresponding
 72-parameter family of vectors defined in~(\ref{GarDNAlgK1})--(\ref{GarDNAdnk})
 are defined from an inner product in the usual way:
 \be\eqalign{
 \angle [  ({\cal R})\, ,({\cal R}^{\prime})]_{\k}
 ~&\equiv~ \cos^{-1} \left( \frac{\left[  ({\cal R}) \, {\bm {\cdot}} \,  ({\cal R}^{\prime}) \right]_{\k}}{{\Big|} ({\cal R}){\Big|}_{\k}{\Big|} ({\cal R}^{\prime}){\Big|}_{\k}} \right) \cr
  \angle [  ({\cal R})\, ,({\cal R}^{\prime})]_{\ell}
 ~&\equiv~ \cos^{-1} \left( \frac{\left[  ({\cal R}) \, {\bm {\cdot}} \,  ({\cal R}^{\prime}) \right]_{\ell}}{{\Big|} ({\cal R}){\Big|}_{\ell}{\Big|} ({\cal R}^{\prime}){\Big|}_{\ell}} \right) 
 }
 \ee
   We find for the length and angles between
 these `vectors' the results:
\be \eqalign{
 & {{\Big|} (CM){\Big|}_{\ell}}^2 ~=~ {{\Big|} (CM){\Big|}_{\k}}^2 ~=~  {{\Big|} (VM)]  {\Big|}_{\ell}}^2 ~=~  {{\Big|} (VM)]  {\Big|}_{\k}}^2
 ~=~  {{\Big|} (TM)]  {\Big|}_{\ell}}^2 ~=~  {{\Big|} (TM)]  {\Big|}_{\k}}^2  ~=~
 6 ~~~, \cr
&\angle [  (CM)\, ,(VM)]_{\ell}
 ~=~  \angle [  (CM)\, ,(TM)]_{\ell}
 ~=~ \fracm {\p}2 ~~~, ~~~ \angle [  (VM)\, ,(TM)]_{\ell}
 ~=~ \cos^{-1}\left[-\, {\fracm 13} \right], \cr
&\angle [  (CM)\, ,(VM)]_{\k}
 ~=~  \angle [  (CM)\, ,(TM)]_{\k}
 ~=~ \fracm {\p}2 ~~~, ~~~ \angle [  (VM)\, ,(TM)]_{\k}
 ~=~ \cos^{-1}\left[\, {\fracm 13} \right]   ~~~.
} \ee
 which implies these three representations describe a space
 of monoclinic symmetry among the Bravais
 lattices as depicted in Fig.~\ref{f:BoobTubeThetan}.
 
Now that the physicist inside the cave can construct a set of $\g$-matrices that
are identical to those which provided the starting point of the physicist in the
$(1,3)$-dimensional spacetime, it is possible to write the massless Dirac equation for
the spinors in the cave in the form
\be
\d{}^{\hi}{}^\hk \, \partial_{\tau} \,\Psi_\hk  ~=~ 0 ~~\to~~  (C \g^{\m}){}^{\hi}{}^\hk 
\partial_{\m} \,\Psi_\hk  ~=~ 0 ~~~,
 \label{DiracE}
\ee
where the $\g$-matrices are constructed solely from the data in (\ref{chiD0F}),
(\ref{V1D0E}), (\ref{tenD0F}),
and (\ref{GarDNAdnk}). The function $\Psi_\hk $ is now permitted to depend on all the four 
coordinates.  In the final expression the $C$-matrix is the spinor metric (see 
appendix A in \cite{G1}).  In other words, the three valise adinkras in Figure 1, Figure 2, and
Figure 3 contain sufficient information to reconstruct the (1, 3) spinor bundle.

Once the full set of four Dirac matrices has been identified as in~(\ref{DiracE}),
the entire Dirac algebra, $\{\Ione,\g^\m,\g^{[\m\n]},\g^{[\m\n\ro]},\g^5\}$ is
obtained ``for free.'' This contains the six matrices $\g^{[\m\n]}$ that generate
the full Lorentz group $\textit{Spin}(1,3)$ in the standard way\cite{L+M,Hall}. That is, the
identification~(\ref{DiracE}) allows the physicist from Plato's Cave to
reconstruct the Lorentz symmetry of the objects casting the shadows that she 
has been examining.

\section{Conclusions}

$~~~~$ In this work, we have relied the results in \cite{G1} to derive the
$(1,3)$-dimensional Dirac operator by using data that are available to a
physicist who is investigating one dimensional representations of 1D, $N=4$
SUSY.  Using data which follow from adinkras, we have given a demonstration that 
if one possesses a description of the one dimensional multiplet that corresponds 
to the $(1,3)$-dimensional chiral, vector and tensor supermultiplets, this is sufficient 
to reconstruct the Dirac equations for the higher dimensional spinors.  There 
are a number of steps that must be demonstrated to show how to use the 
adinkra-based data to reconstruct the appropriate $(1,3)$ Lorentz covariant 
equations for the bosons.  This will be undertaken in a later work.

When contrasting this work with discussion given in  \cite{dimEnhanc},
we see that there is a {\em {large}} amount of data for 1D valise supermultiplets
(described by equation (\ref{chiD0J})) about the spinor representations that are 
related by dimensional extension.  These works do not take advantage
of this fact and it seems likely that any algorithm for constructing higher
dimensional and more complicated SUSY representations might benefit
from a thorough study of the structure described in this work {\rm {prior}}
to applying node lifting.   Related to such efforts, it seems reasonable
to bring to bare the ``No Two-Color Ambidextrous Bow-Tie'' Theorem
discovered in \cite{Bowtie}.  Since all four dimensional SUSY theories 
must also possess consistent two dimensional SUSY truncations, combining
the present results with the ``No Two-Color Ambidextrous Bow-Tie'' Theorem
will likely enhance the efficacy of search algorithms with the goal of the
discovery of new off-shell SUSY representations.

Let us note that although the formula of (\ref{GarDNAdnk}) was derived in the context of
some 4D, ${\cal N}=1$ supermultiplets, it can also be viewed as a definition
of how to define adinkras {\rm {outside}} the context of supersymmetrical theories.
In particular, whenever there are spin-1/2 fermions present in some construction,
this equation defines adinkras (via the L-matrices) that are associated with the 
$\gamma$-matrices that act on the spinors.  In other words, given a set of
$\gamma$-matrices one can use (\ref{GarDNAdnk}) to {\em {derive}} an associated set
of adinkras and  L-matrices without the presence of supersymmetry.  But
of course such adinkras cannot be related to any bosons in such a theory
as this seems only to occur in the presence of supersymmetry. 

There are indications that there is (much) more to this story,
and we will continue our investigations.

 \vspace{.05in}
 \begin{center}
 \parbox{92mm}{{\it ``Suppose further," Socrates says, ``that the man was
  compelled to look at the fire: wouldn't he be struck blind and 
  try to turn his gaze back toward the shadows, as toward what he 
  can see clearly and hold to 
  be real?''}~---~Plato}
 \end{center}

 \newpage 
\bigskip\bigskip
\paragraph{\bfseries Acknowledgments:}
SJG's and KS's research was supported in part by the endowment of the John S.~Toll 
Professorship, the University of Maryland Center for String \& Particle Theory, National
Science Foundation Grant PHY-09-68854.  TH is grateful to the 
Physics Department of the Faculty of Natural Sciences of the University of Novi Sad, 
Serbia, for recurring hospitality and resources.  Some Adinkras were drawn with the help of the \textsl{Adinkramat}~\copyright\,2008 by G.~Landweber.

$$~~$$
\noindent
\appendix

$$~~~$$


\begin{thebibliography}{99}


\bibitem{G1}
S. J. Gates Jr., J. Gonzales, B. MacGregor, J. Parker,
R. Polo-Sherk, V. G. J. Rodgers, and L. Wassink,
``4D {\cal N} = 1 Supersymmetry Genomics (I),''
JHEP {\bf{12}} (2009) 008, [hep{}-th/0902.3830v4].

\bibitem{GRana}
S.\ J.\ Gates and L. Rana, Phys.\ Lett.\ {\bf {B352}} (1995) 50, arXiv [hep-th:9504025];
ibid.\ Phys.\ Lett.\ {\bf {B369}} (1996) 262, arXiv [hep-th:9510151]; S.\ J.\ Gates, Jr., 
W.\ D.\ Linch, III, J.\ Phillips and L.\ Rana, Grav.\ Cosmol.\ {\bf 8} (2002) 96, arXiv 
[hep-th/0109109]. 

\bibitem{ENUF}
S.\ J. Gates, Jr., W.\ D.\ Linch, III, J. Phillips , ``When Superspace Is Not Enough,'' 
Univ. of Md Preprint \# UMDEPP-02-054, Caltech Preprint \# CALT-68-2387,
arXiv [hep-th:0211034], unpublished.

\bibitem{adinkra1}
M. Faux, S. J. Gates Jr., ``Adinkras: A Graphical Technology for Supersymmetric Representation 
Theory,'' Phys. Rev. {\bf{D71}} (2005) 065002, [hep{}-th/0408004v1];

\bibitem{Adnktop}
C.\ F.\ Doran, M.\ G.\ Faux, S.\ J.\ Gates, Jr., T.\ Hubsch, K.\ M.\ Iga, G.\ D.\ Landweber, and
R.\ L.\ Miller,  ``Topology Types of Adinkras and the Corresponding Representations of N-Extended
 Supersymmetry,''   Univ.\ of Maryland preprint UMDEPP-08-010, State University of New York -
 Oneonta SUNY-O-667 e-Print: arXiv:0806.0050; unpublished.

\bibitem{Codes}
C.\ F.\ Doran, M.\ G.\ Faux, S.\ J.\ Gates, Jr., T.\ Hubsch, K.\ M.\ Iga, and G.\ D.\ Landweber,
``Relating Doubly-Even Error-Correcting Codes, Graphs, and Irreducible Representations of
N-Extended Supersymmetry,''
  in ``Discrete and Computational Mathematics,'' p.~53--71, eds.\ F.\ Liu et al.,
 (Nova Science Pub., Inc., Hauppage, 2008), e-Print: arXiv:0806.0051 [hep-th].;
  C.\ F.\ Doran, M.\ G.\ Faux, S.\ J.\ Gates, Jr., T.\ Hubsch, K.\ M.\ Iga, G.\ D.\ Landweber, and
R.\ L.\ Miller,  ``Codes and Supersymmetry in One Dimension,''   Univ.\ of Maryland preprint 
UMDEPP-008-010, State University of New York - Oneonta SUNY-O-667 e-Print: 
arXiv:1108.4124 [hep-th], to appear in Adv.\ in Theor.\ and Math.\ Phys.\ {\bf {15.6}}
;  S.\ J.\ Gates, Jr., J.\ Hallett,  T.\ Hubsch, and 
K.\ Stiffler, ``The Real Anatomy of Complex Linear SuperÞelds,''   Univ.\ of Maryland preprint 
UMDEPP-012-003, e-Print: arXiv:1202.4418 [hep-th], to appear in the Int. Journ. of Mod.
Phys. 

\bibitem{Posets}
Yan X.\ Zhang, ``Adinkras for Mathematicians,'' M.\ I.\ T.\ mathematics department
(2011) e-Print: arXiv:1111.6055  [math-CO], unpublished.

 \bibitem{dimEnhanc}
 M.\ G.\ Faux, K.\ M.\ Iga, and G.\ D.\ Landweber,  ``Dimensional Enhancement via Supersymmetry,'' 
 Adv.\ Math.\ Phys.\  {\bf {2011}} (2011) 259089, e-Print: arXiv:0907.3605 [hep-th];  M.\ G.\ Faux, 
  K.\ M.\ Iga, and G.\ D.\ Landweber,  ``Spin Holography via Dimensional Enhancement,'' 
  Phys.\ Lett.\ {\bf {B681}} (2009) 161-165, e-Print: arXiv:0907.4543 [hep-th];
 T.\ Hubsch, ``Weaving Worldsheet Supermultiplets from the Worldlines Within,''
  e-Print: arXiv:1104.3135 [hep-th].
 
 \bibitem{Bowtie}
 S.\ J.\ Gates and T.\ Hubsch, ``On Dimensional Extension of Supersymmetry: From 
 Worldlines to Worldsheets,'' Univ.\ of Maryland preprint  UMDEPP-011-005, MIT
 Preprint MIT-CTP-4232, e-Print: arXiv:1104.0722 [hep-th], submitted to Adv.\ in 
 Theor.\ and Math.\ Phys. 
  

\bibitem{PCave}
See the webpage at
http://en.wikipedia.org/wiki/Allegory${}_-$of${}_-$the${}_-$Cave
for a discussion of the `Allegory of the Cave.' 

\bibitem{L+M}
 H.\ B.\ Lawson, Jr., M.-L.\ Michelsohn, ``Spin geometry,'' Vol.\ 38 of Princeton Mathematical Series, Princeton University Press, Princeton, NJ, 1989.

\bibitem{Hall} B.\ C.\ Hall, Lie Groups, ``Lie Algebras, and Representations,'' Springer-Verlag, 2003.

\end{thebibliography}
\end{document}